# Why does silicon have an indirect band gap?

Emily Oliphant[1], Veda Mantena,[1] Madison Brod[2], G. Jeffrey Snyder[2], Wenhao Sun[1*]

[1] Department of Materials Science, University of Michigan, Ann Arbor, Michigan 48109, United States
[2] Northwestern University, Materials Science and Engineering Evanston IL, 60208 USA

*Corresponding Author: whsun@umich.edu

**New Concepts**

Electronic band structure is represented in reciprocal space, but arises from the chemical bonds between atoms in real space. Because bonding environments in crystals are so complex, it has been difficult to isolate which specific atomic orbitals contribute to shaping certain band structure features. For this reason, design and discovery of new semiconductors typically proceeds through a 'needle-in-a-haystack' approach, where high-throughput DFT screening approaches are guided by simple and limited chemical heuristics. Here, we present a new conceptual and computable framework to extract chemical bonding insights from DFT-calculated band structures, enabling us to rigorously and intuitively trace the impact of individual crystal bonds on band structure features. As a key example, we explain here the low-symmetry conduction band minimum of silicon, which profoundly impacts its properties for use in photovoltaics and electronics. Even in this basic semiconductor material, our approach leads to new insights to understand and engineer its conduction band minimum position. These calculation techniques can be broadly applied to reveal the crystal chemistry origins of electronic structure features in other optical, electronic and magnetic materials.

**Abstract**

It is difficult to intuit how electronic structure features—such as band gap magnitude, location of band extrema, effective masses, *etc*.—arise from the underlying crystal chemistry of a material. Here we present a strategy to distill sparse and chemically-interpretable tight-binding models from density functional theory calculations, enabling us to interpret how multiple orbital interactions in a 3D crystal conspire to shape the overall band structure. Applying this process to silicon, we show that its indirect gap arises from a competition between first and second nearest-neighbor bonds—where second nearest-neighbor interactions pull the conduction band down from Γ to X in a cosine shape, but the first nearest-neighbor bonds push the band up near X, resulting in the characteristic dip of the silicon conduction band. By identifying the essential orbital interactions that shape the conduction band, we can further rationally tune bond strengths to morph the silicon band structure into the germanium band structure. Our computational approach serves as a general framework to extract the crystal chemistry origins of electronic structure features from density functional theory calculations, enabling a new paradigm of *bonding-by-design*.

**Introduction**

Silicon has an indirect band gap, with the valence band maximum (VBM) at the Γ point and the conduction band minimum (CBM) at a low-symmetry point ~85% of the way between the Γ and X points. This indirect band gap determines the essential electronic and optical properties of silicon, and thereby its performance in photovoltaic and electronic devices.[1–3] Although the low-symmetry CBM of silicon is a basic fact of semiconductor physics, it is not so simple to answer *why* silicon has an indirect band gap. Similar 'why' questions can be generally raised about the electronic structures of materials. Why does germanium have a CBM at the L point, despite also being a Group IV semiconductor in the diamond structure? Why does zincblende GaAs have a direct band gap, with such a light electron effective mass? Without a conceptual framework to approach *why* questions, one must rely on simple heuristics, which may post-rationalize the chemical origins of band structure but fall short in accurately predicting band structure features. Consequently, the search for next-generation thermoelectrics, *p*-type transparent conducting oxides, topological insulators, and other advanced electronic materials[4–6] must proceed by brute-force screening via a "needle-in-a-haystack" approach [7–9], rather than by rational and intuitive design.

Roald Hoffman presented a beautiful theoretical framework to examine how physics and chemistry meet in the solid-state[10–12], arguing that chemists approach electronic structure from a bottom-up Linear Combination of Atomic Orbitals (LCAO). Alternatively, physicists adopt a top-down planewave interpretation of electronic structure, often using density functional theory (DFT) for accurate band structure calculations. The tight-binding (TB) model is a periodic version of LCAO that offers a bridge from bonds to bands, and back again.[13] If a band in the electronic structure is dominated by a single orbital interaction, tight-binding offers a conceptual pathway to interpret how bonding in the wavefunction modulates across reciprocal space, thus explaining the band's shape. However, when bands in 3D materials are formed by multiple orbital interactions, from multiple neighboring atoms, it becomes difficult to deconvolute how specific orbitals conspire to shape a band structure feature.

Given this complexity, creating tight-binding (TB) models in 3D structures requires a fundamental trade-off between simplicity and interpretability. Simple tight-binding models are usually constrained to first nearest neighbors (1NN), where one asserts *a priori* which bonds are considered. While the resulting models are usually easy to interpret, they may not be physically robust. In the case of silicon, achieving a CBM near the X point using 1NN tight-binding has relied on including an additional *s\** and/or *d* states, which have limited physical relevance as they may not describe the precise physics of the actual excited states.[14–22] On the other hand, tight-binding models that consider further atomic neighbors, such as by interpolation from DFT band structures or fitting with many-NN, increases the accuracy of a TB model but the model combinatorically explodes in the number of terms—precluding chemical interpretability[23–35].

This paper aims to bridge the chemical intuition of Hoffmann and the practical toolkit of DFT, so that we can better interpret the chemical origins of electronic structure in real materials. To do so, we start from a tight-binding interpolation of the DFT-calculated electronic structure using maximally localized Wannier functions (MLWFs).[36] From this tight-binding interpolation, we chemically interpret how band shapes in *k*-space derive from orbital interactions in real space. We present a three-step process, illustrated in **Figure 1**, which proceeds by determining: **1)** *Which orbitals contribute to a band*—specifically, what are the orbital characters (coefficients) that contribute to the wavefunction. **2)** *How orbitals bond across* k-*space*—where the *k*-dependent phase ($e^{ik \cdot R}$) of each orbital changes the bonding/antibonding/non-bonding interactions between the orbitals in real space. **3)** *How strongly the orbitals bond*—determined by the magnitudes of the TB hopping parameters. These three steps systematically sieve through hundreds of TB parameters to build a sparse and physically robust model to interpret chemical bonding contributions to the electronic structure.

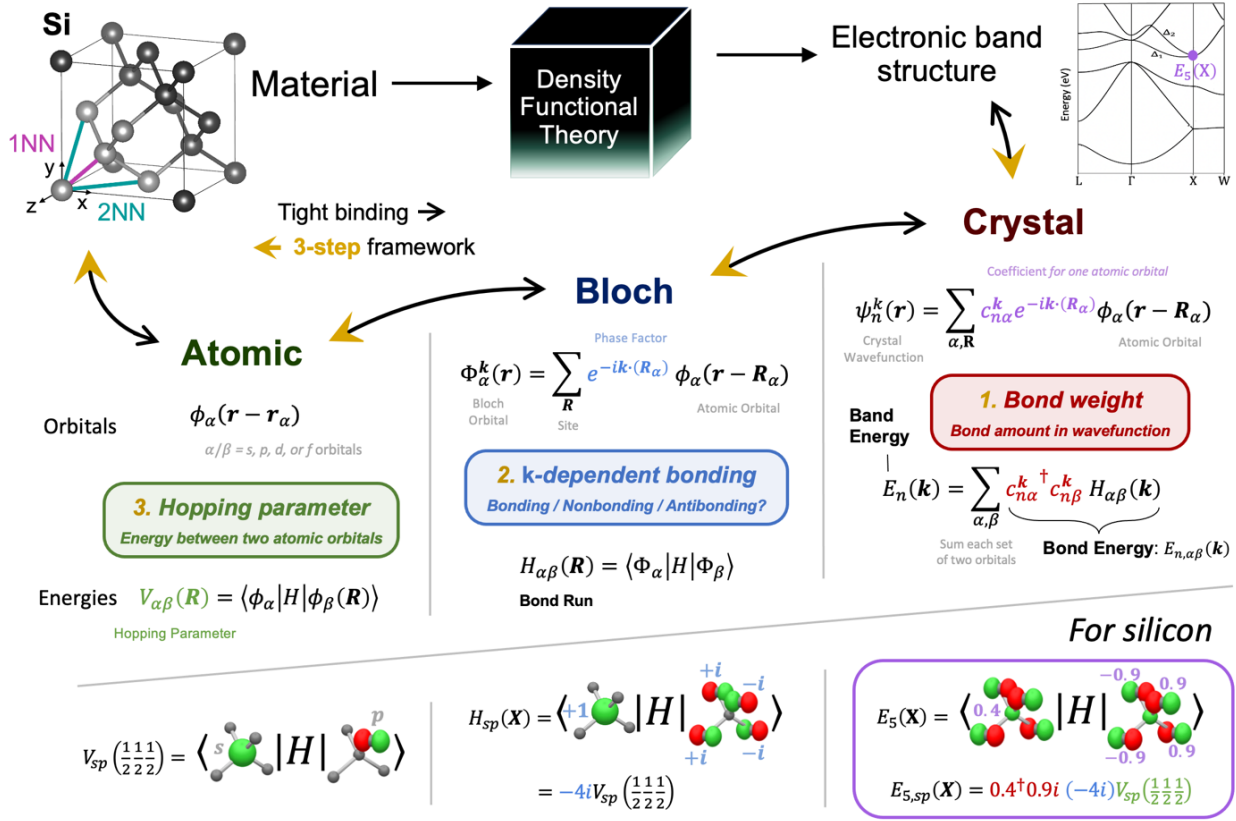

*Figure 1: Tight-binding offers a pathway from crystal chemistry to band structure. Our 3-step framework explains the reverse path—how to interpret a DFT electronic structure with the chemical understanding of tight binding. In the 'For silicon' section bright green and red is used for positive and negative isosurfaces for example atomic orbitals, Bloch orbitals, and crystal wavefunctions.*

Here, we begin with an illustrative 1D model system to emphasize the impact of multi-orbital and >1NN interactions on band structure. Then we apply our three-step process to build a concise and chemically robust TB interpretation for how multiple orbital interactions combine to form a low-symmetry CBM in Silicon. Specifically, we find that this low-symmetry CBM along the Γ-X line derives primarily from a cosine shape from second nearest neighbor (2NN) $p_x$-$p_x$ bonds, combined with a linear shape near X from 1NN orbital bonds. Finally, we present a new computational tool to interactively execute our three-step process for any band and *k*-point in an electronic band structure (https://viz.whsunresearch.group/tb/). This tool enables us to precisely identify which bonds affect which segments of the silicon band structure. We then rationally tune these orbital interactions to morph the silicon band structure towards the germanium band structure, in accordance with the actual chemical differences between Si and Ge. Altogether, our work serves as a general blueprint to extract the crystal chemistry origins of electronic band structure, and provides a pathway for rational band-structure engineering by chemical and structural design.

**Uncovering the chemistry in band structure**

All electronic properties of a material—including band gap, effective mass, band extrema location, *etc.*—are characterized by its band structure. The chemical origin of these electronic properties can be elucidated from accurate TB interpolations of DFT-calculated electronic structure. Tight-binding decomposes a band structure into a summation of terms, similar to a Fourier decomposition, but with basis functions that correspond to physically-relevant orbital interactions.[37] This decomposition enables one to trace back which bonds manifest which specific band features of interest. **Figure 1** offers a hierarchical mind-map that captures the tight-binding pathway from crystal chemistry to the band structure.

The central assumption of tight binding is to write the crystal wavefunction, $\psi_n^k$, as a linear combination of atomic Bloch orbitals, $\Phi_\alpha^k$, weighted by the coefficient $c_{n\alpha}^k$ as in **Equation 1**. The indices $k$, $n$, and $\alpha$ represent $k$-point, band, and orbital. An atomic Bloch orbital is the sum over atomic orbitals, $\phi_\alpha$, in each cell of the crystal related by a lattice vector translation **R**. The complex phase of each atomic orbital is modified by the phase factor $e^{-i\boldsymbol{k}\cdot(\boldsymbol{R}_\alpha)}$, where $\boldsymbol{R}_\alpha = \boldsymbol{r}_\alpha + \boldsymbol{R}$ and $\boldsymbol{r}_\alpha$ is the atomic orbital center in the primitive cell.

$$|\psi_n^k\rangle = \sum_\alpha c_{n\alpha}^k |\Phi_\alpha^k\rangle = \sum_{\alpha,R} c_{n\alpha}^k\, e^{-i\boldsymbol{k}\cdot(\boldsymbol{R}_\alpha)}\, |\phi_\alpha(\boldsymbol{R}_\alpha)\rangle \qquad \text{Eqn. 1}$$

With an atomic decomposition of the crystal wavefunction, each band dispersion (shape in $E$-$k$ space) expands as the combination of numerous pairwise bonds between atomic orbitals. In **Equation 2**, the overall shape of the band, $E_n(\boldsymbol{k})$ for band $n$, results from the sum of 'bond energies' $E_{n,\alpha\beta}(\boldsymbol{k})$—which is the $E$-$k$ shape of the bond between Bloch orbitals $\alpha$ and $\beta$.

$$E_n(\boldsymbol{k}) = \langle \psi_n^k | \hat{H} | \psi_n^k \rangle = \sum_\alpha \sum_\beta {c_{n\alpha}^k}^\dagger c_{n\beta}^k \langle \Phi_\alpha^k | \hat{H} | \Phi_\beta^k \rangle = \sum_\alpha \sum_\beta E_{n,\alpha\beta}(\boldsymbol{k}) \qquad \text{Eqn. 2}$$

This equation substitutes the crystal wavefunction with the middle expression in **Eqn. 1**. From this, the bond energy, $E_{n,\alpha\beta}(\boldsymbol{k})$, is the bond weight, ${c_{n\alpha}^k}^\dagger c_{n\beta}^k$, multiplied by the *bond run*, which we name the Hamiltonian matrix element between Bloch orbitals $\alpha$ and $\beta$, $\langle \Phi_\alpha^k | \hat{H} | \Phi_\beta^k \rangle$. The name 'bond run' is inspired by Hoffmann's discussion[10] that bands made from *s* orbital bonds 'run up' in energy from Γ to the Brillouin zone edge (from bonding to antibonding), while bands from *p* orbital bonds 'run down' from Γ to the Brillouin zone edge (from antibonding to bonding).

To isolate the impact of atomic orbital bonds, the bond run—written as $H_{\alpha\beta}(\boldsymbol{k})$—is expanded by substituting the Bloch orbitals with the sum of $k$-modulated atomic orbitals, as in **Eqn. 1**. This results in **Equation 3**, which is a sum over bonds between an atomic orbital $\alpha$ at $\boldsymbol{r}_\alpha$ and atomic orbital $\beta$ at $\boldsymbol{r}_\beta + \boldsymbol{R}$. Each interaction is then the TB hopping parameter, $\langle \phi_\alpha | \hat{H} | \phi_\beta(\boldsymbol{R}) \rangle \equiv V_{\alpha\beta}(\boldsymbol{R})$, multiplied by a factor which modulates the complex phase based on the phase difference between orbital centers for a given $k$.

$$\langle \Phi_\alpha^k | \hat{H} | \Phi_\beta^k \rangle \equiv H_{\alpha\beta}(\boldsymbol{k}) = \sum_R V_{\alpha\beta}(\boldsymbol{R}) e^{i\boldsymbol{k}\cdot(\boldsymbol{r}_\alpha - \boldsymbol{r}_\beta - \boldsymbol{R})} \qquad \text{Eqn. 3}$$

Here, the $k$-dependence of a bond run arises from the sum of phase factors $e^{i\boldsymbol{k}\cdot(\boldsymbol{r}_\alpha - \boldsymbol{r}_\beta - \boldsymbol{R})}$. When the bond run is negative for a given $k$, it indicates the Bloch orbitals $\alpha$ and $\beta$ are overall bonding, whereas positive indicates net antibonding. While the phase of the coefficients ${c_{n\alpha}^k}^\dagger c_{n\beta}^k$ must also be included to determine the precious bond type for each band, this complication is reserved for later when Eqn. 4 is introduced.

Our three-step process is grounded in these two fundamental equations, offering a way to back solve from a given band structure the chemical bonding contributions, as diagrammed in **Figure 1**. **Step 1** is bond weight, where ${c_{n\alpha}^{k}}^{\dagger} c_{n\beta}^{k}$ quantifies the bonds (between orbitals $\alpha$ and $\beta$) that may contribute to the band. While **Step 2** is the shape of the bond run, which describes how the bond-type (bonding, antibonding, or non-bonding) changes across $k$-space. **Step 3** is the maximum absolute energy of the bond, dictated by the hopping parameter between two atomic orbitals $\alpha$ and $\beta$, $V_{\alpha\beta}(\mathbf{R})$.

**Eqn. 2** is the same theoretical starting point as the crystal orbital Hamiltonian population [38,39] method, but the implementation of an analytical representation in **Eqn. 3** (instead of a numerical calculation, for example as done through LOBSTER[40]) requires a tight-binding interpolation, which enables us to then break down the contributions from individual bonds. These analytical representations of each orbital wavefunction lets us visualize and further separate into NN or long-range interactions.

To examine how multiple orbitals and further nearest-neighbor interactions manifest in both real space wavefunction and the reciprocal space bands, here we present an illustrative example on a one-dimensional monatomic chain, with two orbitals ($s$ and $p$) per atom. Traditionally, 2NN and further terms are neglected in simple TB models, but here we show how they could induce major qualitative changes to the band structure. The details of the derivation are in **SI.1**.

Plotted in **Figure 2a**, the 1NN bond runs $H_{ss}$ and $H_{pp}$ are cosine curves with extrema at the high symmetry points, but all other bond runs ($H_{sp}$ and 2NN bonds) have extrema at low symmetry points. Using **Eqn. 2**, we sum together the bond runs multiplied by their bond weight (orbital character) to find the band energy. With only 1NN, $E_2(k)$ is dominated by $H_{pp}(k)$, which has extrema at high symmetry points. We next add 2NN interactions, with an exaggerated bond strength of half the 1NN, which results in a qualitative change in the overall band shape. This additional strong 2NN interaction pulls the bands together near $\frac{X}{2}$ while pushing them apart near $\Gamma$ and $X$, creating a low-symmetry band extremum reminiscent of silicon.

When we ground our tight-binding intuition in 1D models, the 2NNs are so far away that their contributions tend to be small. However, in 3D crystals, atoms have much higher coordination numbers. These 2NNs are also much closer in 3D crystals than they are in 1D systems, meaning the 2NN contribution to the tight-binding interactions can be substantial. In the case of silicon, the twelve 2NNs are only 1.6× further than the four 1NNs, with a geometry that allows for strong overlap between $p$ orbitals. By studying silicon, we will show how bonds of similar strength with different frequency of bond runs leads to band extrema away from high symmetry points. Another common cause for low symmetry band extrema is an avoided crossing from $s$-$p$ orbital mixing. However, avoided crossings will always result in a band inversion (switching of orbital character), which the silicon CBM does not exhibit. Thus, if there is a low symmetry band extremum on a band that does not have a band inversion, long-range interactions beyond 1NN are a likely culprit.

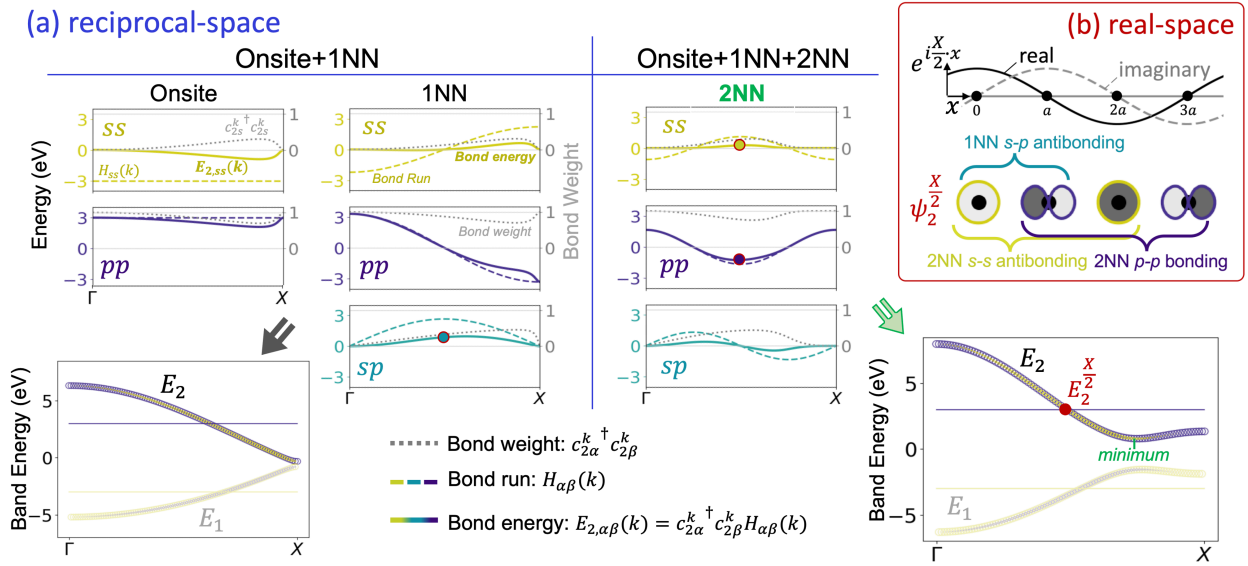

*Figure 2: Multiple bonds combine to form band energy with a monoatomic s+p 1D tight-binding model. When adding 2NNs, band extrema at low-symmetry k-points manifest.* In (a), the bond runs, bond weights[41], and bond energies are plotted for onsite (orbital energy), 1NN, and 2NN bond contributions to the second band, $E_2$. The bond energies sum to create the $E_2$ band energy, where the orange/blue circle size indicates orbital character, $|c_{s/p}^k|$. To make them real, $-iH_{sp}(\mathbf{k})$ and $ic_{2s}^{k\dagger}c_{2p}^k$ are plotted for the sp bonds. On the right, 2NN terms are added, perturbing the bond weights and band energies to create a low symmetry extremum off-X. In (b), the real part of the $\Psi_2^{X/2}$ wavefunction is plotted to highlight the real-space bonding implicit in band structure. The orbital at each atom is determined by the phase factor, $e^{i(X/2) \cdot x}$, multiplied by the orbital coefficients, $c_s = 0.49$ and $c_p = -0.87i$. The nonzero bond energies are written and circled in red on the reciprocal-space plots.

In our ambition to eventually design band structure from the underlying bonds, we need to first elucidate the orbital and bonding nature at the specific *k*-point of a band. The key term linking real space chemical bonding with reciprocal space band structure is the phase factor $e^{i\mathbf{k}\cdot\mathbf{R}}$, manifesting in the crystal wavefunction of **Eqn. 1**. For a given *k*-vector, the combined coefficient $e^{i\mathbf{k}\cdot\mathbf{R}_\alpha} c_{n\alpha}^k$ for each atomic orbital $\alpha$ dictates the bond type (bonding, antibonding, or non-bonding) between atomic orbitals in real space **(Step 2)**. If the complex phases of neighboring orbitals are orthogonal (e.g. real and imaginary), they do not interact and are non-bonding. If they are nonorthogonal (e.g. real and real, or imaginary and imaginary), they are bonding or antibonding, depending on the signs of the wavefunction.

In **Figure 2b**, we illustrate the relationship between phase factor and real-space $\psi_2^{X/2}$ wavefunction at *k* = X/2, which requires a 4-unit cell superstructure in real space. Across the 4 atoms in **Figure 2b**, the phase factor $e^{i(X/2)\cdot R}$ modulates as +1, +*i*, –1, and –*i*. Because $c_2^k$ coefficients are real and positive for *s* orbitals, whereas they are imaginary and negative for *p* orbitals, we see that the real part of $\psi_2^{X/2}$ has *s* orbitals on atoms 1 and 3; and *p* orbitals on atoms 2 and 4. Therefore, the only 1NN interaction is *s-p* antibonding. The phase sign switching between atoms 1 (2) and 3 (4) yields 2NN *s-s* antibonding (2NN *p-p* bonding). This example visualizes how chemical bonding in real space implicitly derives from each *k*-point in reciprocal space.

**Detangling the Silicon band structure**

Although silicon has been studied for decades, the crystal chemistry origins of its low-symmetry conduction band minimum still lack satisfactory explanation. Tight-binding (TB) models fitted with only 1NN incorrectly produce a conduction band minimum at Γ. While Vogl produced a CBM off-X with an additional excited $s^*$ state, he acknowledges that "The inclusion of some such excited states in any minimal basis set is physically important–although the precise physics of the actual excited states need not be faithfully and quantitatively reproduced."[16] Indeed, while the $sps^*$ model fits the Γ-X line, it sacrifices the accuracy of the rest of the conduction band structure along nearly every other $k$-path, as detailed in **SI.2**.

Since band structure arises from the complex interactions between multiple orbitals, it is often possible to have multiple non-unique solutions that fit a singular band feature. Following Vogl, others have included additional orbital states to their TB models—for example Jancu et. al. and others added $d$ states, producing a reasonable fit of the lowest conduction bands but again with little physical insight regarding the additional parameters.[17–22] Tight-binding models with >1NN were also fit (often with $s^*$ states) for silicon and zinc-blende semiconductors with varying degrees of 2NN contribution.[23–32,35]

From a model-building perspective, it is not satisfactory to include terms *ad hoc* just to match a single band structure feature—rather, a term that is physically valid should improve the fit of all band energies throughout the entire Brillouin Zone. This is especially important if one aims to later engineer and design the band structure by modifying chemical interactions, which requires one to accurately identify the true chemical origin of band features.

At the other end of the spectrum, one can perform a TB interpolation directly from DFT—which obtains the hopping parameters from a Fourier transformation of the $k$-dependent orbital Hamiltonian. This TB interpolation circumvents the need to assume which interactions are present, but the resulting many-NN TB models can have hundreds of non-trivial interactions, which is too complex to interpret chemically. Sanchez-Portal, and later Qian et al, applied a TB interpolation which includes many NNs to silicon finding a low symmetry minimum along the Γ-X line with only a $sp$ basis, indicating that $s^*$ and $d$ states are not strictly required to reproduce the minimum off X.[33,42] Since then, TB interpolations of silicon are frequently achieved using MLWFs and similar methods, but a simple chemical understanding has not yet been detangled from the hundreds of hopping parameters found.

Here, we apply our three-step process to build a chemical interpretation for the conduction band minimum in silicon along the $\Delta_1$ band. Our DFT calculations were done with the Vienna Ab initio Simulation Package using Perdew-Burke-Ernzerhof pseudopotentials, a plane wave energy cutoff of 520 eV, and $k$-point density of 0.23 Å$^{-1}$.[43–45] Details of our MLWF parameters and process are discussed in **SI.3**. Our 3-step process is implemented as follows: **Step 1)** We determine which orbitals contribute to the $\Delta_1$ band—finding that it is >80% $p_x$ orbital character, with the remaining character being $s$ orbitals. **Step 2)** we determine how the orbitals bond across $k$-space—finding from the bond runs that the second nearest neighbor $p_x$-$p_x$ is the only interaction that decreases the band energy at the X point. **Step 3)** we determine how strongly the orbitals bond—showing that the 2NN $p_x$–$p_x$ bond has large hopping parameters and high coordination which makes it a significant influence on the band structure.

Finally, individual chemical bonds are assessed for their contribution to the total shape of the $\Delta_1$ band. From this, we determine that the low-symmetry conduction band minimum of silicon manifests from a linear shape of the 1NNs near X, combined with the cosine shape of the 2NN $p_x$–$p_x$ bond. Crucially, including the 2NN $p_x$–$p_x$ bond not only improves the $\Delta_1$ band, but the band structure across all other high-symmetry lines (details in **SI.2**), validating its physical significance in creating the low-symmetry Si CBM.

## 1) *Orbital character of bands*

First, we determine which orbitals in the $\Delta_1$ band are present to bond. In a *sp* model, silicon in the diamond structure has eight orbitals, four for each of the two atoms in the primitive cell. This amounts to 72 Hamiltonian matrix elements—8 onsite, 32 1NN, and 32 2NN interactions. After symmetry and group theory considerations, a wavefunction along the Γ–X line will have either $s + p_x$ orbitals or $p_y + p_z$ orbitals. With only $s$ and $p_x$ orbitals are on the $\Delta_1$ band, the 72 matrix elements can be reduced to 8 unique elements. To separate the character of a general band, it is essential to use atomic orbitals as the momentum-dependent crystal wavefunctions rarely reduce to the hybrid atomic orbitals of simple molecular wavefunctions.

**Figure 3** plots positive (red) and negative (green) isosurfaces for the real part of the complex wavefunctions in one of each of the four doubly-degenerate bands at X. For all the $X_1$ bands, the first atom in the basis set has only $p_x$ orbitals and the second atom has only (distorted) $s$ orbitals. Away from X, the $X_1$ degeneracy splits into $\Delta_1$ (CBM band) and $\Delta_2$ bands with $s$ and $p_x$ orbital character, while the $X_4$ bands remain degenerate. In Silicon, tight-binding analysis of the $\Delta_1$ conduction band character shows it is predominantly (>80%) $p_x$ orbital character. The $X_1$ antibonding wavefunction is mainly 1NN $s$-$p_x$ antibonding and 2NN $p_x$-$p_x$ bonding. Importantly, the 2NN $p_x$–$p_x$ interaction is the only one that is bonding along the *x*-direction and contributes to lowering the energy at X.

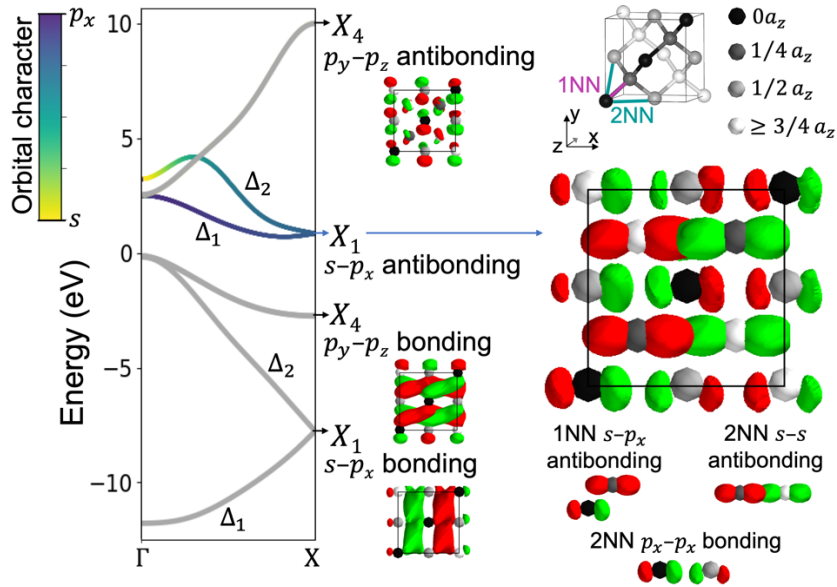

*Figure 3: The crystal wavefunctions at the X point in silicon with the $X_1$ conduction band highlighted to show orbitals and bond type.* Each of the four doubly degenerate bands is accompanied by the present 1NN bond and the plotted real part of a wavefunction. The atom sites are spheres colored to indicate the z-coordinate. The red and green show the positive and negative isosurfaces of the real wavefunction, where neighboring same color lobes are bonding and different color lobes are antibonding. Bonding lobes often mesh together while antibonding lobes are distorted apart. By looking closer at the $X_1$ band, we determine 1NN and 2NN bond-types, where the 2NN $p_x - p_x$ interaction is the only bonding along the x-direction.

Although these arguments explain the lowering of the $\Delta_1$ band energy *at* X, the actual CBM is at a low-symmetry point ~85% of the way from Γ to X. In the next two steps, we examine the shape (Step 2) and magnitude (Step 3) of the participating bonds. Then in the final section, we combine multiple bonds to achieve the $\Delta_1$ band dispersion and observe how the minimum off-X manifests.

## 2) *Shape of bonds in k-space*

Second, we uncover how bonding changes with *k*-space. The real-space bond type can be visualized in the full wavefunctions by mapping the *k*-dependent phase modulations onto the atomic orbitals, as in **Figure 2b**. For the sake of brevity, we have included this visualization and discussion in **SI.4**. In this section, we focus on how the real-space bonding manifests as higher or lower energies in reciprocal space, forming the shape that each bond has along the $\Delta_1$ band.

As discussed in the first section, a bond's contributed shape is referred to as the bond energy, which is the bond run multiplied by bond weight. Because the bond run defined in **Eqn. 3** can be any phase and does not vary with band, here we introduce a band-dependent bond run $H_{n,\alpha\beta}^R(k)$, **Equation 4**. This includes the phase of the orbital coefficients for band $n$, ensuring that $H_{n,\alpha\beta}^R$ is real and the correct sign for band $n$. The superscript R is included to indicate that only the primitive cell vectors for onsite, 1NN, or 2NN bonds are included during the sum over R in **Eqn. 3**. Finally, using **Eqns. 2** and **4** we write the bond energy, $E_{n,\alpha\beta}^R(k)$, as band-dependent bond run multiplied by the absolute value of the orbital coefficients, **Eqn. 5**.

$$H_{n,\alpha\beta}^R(k) \equiv \frac{c_{n\alpha}^{k\,\dagger} c_{n\beta}^k}{|c_{n\alpha}^k||c_{n\beta}^k|} H_{\alpha\beta}^R(k) \qquad \textbf{Eqn. 4}$$

$$E_{n,\alpha\beta}^R(k) = |c_{n\alpha}^k||c_{n\beta}^k| H_{n,\alpha\beta}^R(k) \qquad \textbf{Eqn. 5}$$

In the top of **Figure 3**, we plot the band-dependent bond runs $H_{\Delta_1,\alpha\beta}$ and the bond energies $E_{\Delta_1,\alpha\beta}$ for the onsite, 1NN, and 2NN interactions. The bond run shapes are similar to the illustrative 1D example from **Figure 2a**, but with half the length in reciprocal space since silicon has a two-atom primitive cell. The bond run magnitudes are dictated by the hopping parameter. The onsite orbital energy terms are most simple, for example the onsite $p_x$ term is $E_{\Delta_1,xx}^0 = \varepsilon_p |c_{\Delta_1,x}^k||c_{\Delta_1,x}^k|$, where $\varepsilon_p$ is the $p_x$-like orbital energy, $c_{\Delta_1,x}^k$ is the $p_x$ orbital coefficient. The 1NN and 2NN terms require more derivation, which is left to **SI.5**. Most importantly, the 2NN $p_x$-$p_x$ orbital bond energy is $E_{\Delta_1,xx}^{2NN} = |c_{\Delta_1,x}^k||c_{\Delta_1,x}^k| \cdot [8V_{xx}(110)\cos(k_x\pi) + 4V_{xx}(011)]$, the right part of which is the bond run $H_{\Delta_1,xx}^{2NN}(k)$, which reveals how the 2NN $p_x$-$p_x$ interaction changes bond-type throughout the $\Delta_1$ band. Crucially, the dominant 1NN $s$-$s$ and $s$-$p$ bonds are higher in energy at X than Γ, encouraging a CBM at Γ. Whereas the positive cosine in the bond run for the 2NN $p_x$–$p_x$ interaction lowers the energy at X, enabling a minimum near X.

### 3) *Strength of hopping parameters*

Third, we examine the magnitude of atomic orbital interactions using hopping parameters. Based on intuition from 1D models, we would anticipate the hopping parameters for 1NNs to generally be much larger than for 2NNs. However, here we find that the 2NN bonds are very important in silicon, as supported by the hopping parameters from our atomic-like MLWF tight-binding interpolation in **Table 1**. When including bond multiplicity, the twelve 2NN $p_x$–$p_x$ parameters sum as $8V_{xx}(110) + 4|V_{xx}(011)| = 2.00$ eV, which is 4× larger than the four 1NN $p_x$–$p_x$ parameters $V_{xx}\left(\frac{1}{2}\frac{1}{2}\frac{1}{2}\right)$. Here, $V_{xx}(110)$ indicates the hopping parameter between $p_x$–$p_x$ orbitals on atoms separated by the vector $1\textbf{x} + 1\textbf{y} + 0\textbf{z}$ (or a symmetrically equivalent vector), where the Cartesian $\textbf{xyz}$ unit vectors are half the unit cell length. Combined with the dominant orbital character being $p_x$, the 2NN $p_x$–$p_x$ contributes significantly to the energy of the $\Delta_1$ wavefunctions. Other tight-binding models which have included 2NN parameters either did not include a $V_{xx}(110)$ term, or they were ~10× smaller than our MLWF-derived result.[24–32] An exception to this is Grosso & Piermarocchi who fit a $V_{xx}(110)$ about 2× larger than our result in **Table 1**.[35] In all cases, the 2NN contributions were not individually analyzed for their role in shaping the CBM.

Table I. Silicon hopping parameters (eV) from MLWF used to reconstruct the $\Delta_1$ band. The s and p orbital onsite terms are $\varepsilon_s$ and $\varepsilon_p$. Importantly, the 2NN terms $V_{xx}(110)$ and $V_{xx}(011)$ are similar or larger than the 1NN term $V_{xx}\left(\frac{1}{2}\frac{1}{2}\frac{1}{2}\right)$.

| $\varepsilon_s$ | $\varepsilon_p$ | $V_{ss}\left(\frac{1}{2}\frac{1}{2}\frac{1}{2}\right)$ | $V_{sx}\left(\frac{1}{2}\frac{1}{2}\frac{1}{2}\right)$ | $V_{xx}\left(\frac{1}{2}\frac{1}{2}\frac{1}{2}\right)$ | $V_{xx}(110)$ | $V_{xx}(011)$ |
|---|---|---|---|---|---|---|
| -5.467 | 1.650 | -1.639 | 1.075 | 0.126 | 0.117 | -0.267 |

The small hopping parameter between 1NN $p_x$–$p_x$ orbitals can be understood from the geometry as discussed by Slater.[46] The hopping parameter between two $p_x$ orbitals is given by $l^2 V_{pp\sigma} + (1-l^2)V_{pp\pi}$, where $l$ is the direction cosine in the $x$ direction. If the $p_x$-$p_x$ lobes are facing each other (like ∞-∞) then there is perfect σ bonding, where $l = 1$, whereas if $p_x$-$p_x$ lobes are parallel (like 8-8) then $l = 0$ and there is perfect π bonding. Because $V_{pp\sigma}$ and $V_{pp\pi}$ have opposite sign, an intermediate orientation between perfectly aligned (∞-∞) and perfectly parallel (8-8), will result in the hopping parameter canceling to zero.

In the tetrahedral coordination environment of the diamond structure, 1NN have $l^2 = 1/3$, such that the $V_{pp\sigma}$ and $V_{pp\pi}$ components nearly cancel. The 2NN have four neighbors with $l^2 = 0$, meaning the $V_{xx}(011)$ parameters are entirely π bonding, while the other eight neighbors have $l^2 = 1/2$, allowing the stronger σ antibonding to dominate the weaker π bonding in the $V_{xx}(110)$ parameters.

### Steps 1+2+3: The low-symmetry $\Delta_1$ minimum in silicon

Finally, we create the total shape of the silicon $\Delta_1$ band dispersion from individual bonds and determine which bonds are necessary to capture the correct qualitative band shape. Each bond contributes a distinct shape over some high-symmetry line of the band structure. To obtain the correct band dispersion that matches a DFT band structure, all significant bonds must be included. As Vogl showed, a 1NN tight-binding model with an $sp$ basis cannot produce a low-symmetry conduction band minimum[16], whereas an many-NN $sp$ basis can [33,42]. This indicates that physically significant bonds are missing from the 1NN $sp$ model, which as we have argued, are the 2NN interactions.

The $\Delta_1$ band energy as a function of the bonds can be simplified from the sum over each matrix element, **Eqn. 2**, to a sum over unique elements for the $s$ and $p_x$ (written as $x$) orbitals, **Eqn. 6**. As discussed earlier, this reduces the 72 parameters for the $\Gamma - X$ line to only 8: two onsite (orbital energy) terms, three 1NN terms, and three 2NN terms.

$$E_{\Delta_1}(k) = 2E^0_{\Delta_1,ss} + 2E^0_{\Delta_1,xx} + 2E^{1NN}_{\Delta_1,ss} + 4E^{1NN}_{\Delta_1,sx} + 2E^{1NN}_{\Delta_1,xx} + 2E^{2NN}_{\Delta_1,ss} + 4E^{2NN}_{\Delta_1,sx} + 2E^{2NN}_{\Delta_1,xx} \quad \textbf{Eqn. 6}$$

Each term has an analytical expression as seen from **Eqn. 5**, which is the product of relevant orbital coefficients (a multiplicative factor) with the band-dependent bond run (a cosine or sine shape). For full derivation and decomposition of **Eqn. 6**, see **SI.5**.

To conceptually understand how the CBM arises from multiple orbital interactions, we compare and combine the relevant bond runs and energies in the top panels of **Figure 4** to create the bottom panel. When looking at the bond runs, the onsite and 1NN terms dominate the 2NN, with the 1NN $s$–$p_x$ and $s$–$s$ spanning ~8 eV each, while the 2NN $p_x$–$p_x$ only reaches 2 eV. But once the strong $p_x$ orbital character is included with $E_{\Delta_1,\alpha\beta}$, the $p_x$–$p_x$ terms are nearly unchanged, while the $s$–$p_x$ decrease significantly, and the

$s$–$s$ drops nearly to zero. This puts the [onsite + 1NN] energy magnitude in the same range as the 2NN $p_x$–$p_x$, where each span ~1.5 eV (**Figure 4** bottom panel). The shape of onsite + 1NN is a pseudo-linear increase near X, which results primarily from the $E^{1NN}_{\Delta_1,ss}$ shape. The cosine-like shape of $E^{1NN}_{\Delta_1,ss}$ results from the sine curve of $H^{1NN}_{\Delta_1,ss}$ being heavily distorted by the coefficient weight $|c_{\Delta_1 s}||c_{\Delta_1 p}|$ increasing from $\Gamma$ to X.

Finally, the pseudo-linear shape of the onsite + 1NN near X plus the cosine curve of the 2NN $p_x$–$p_x$ combine to form the conduction band minimum away from the high-symmetry X point in silicon. Importantly, including the 2NN $p_x$–$p_x$ bond also provides a good band structure fit on all other $k$-paths (see **SI.2** for details), compared to the $s^*$ state from Vogl et al.[16], which validates the physical importance of 2NN bonding in governing the low-symmetry conduction band minimum of silicon.

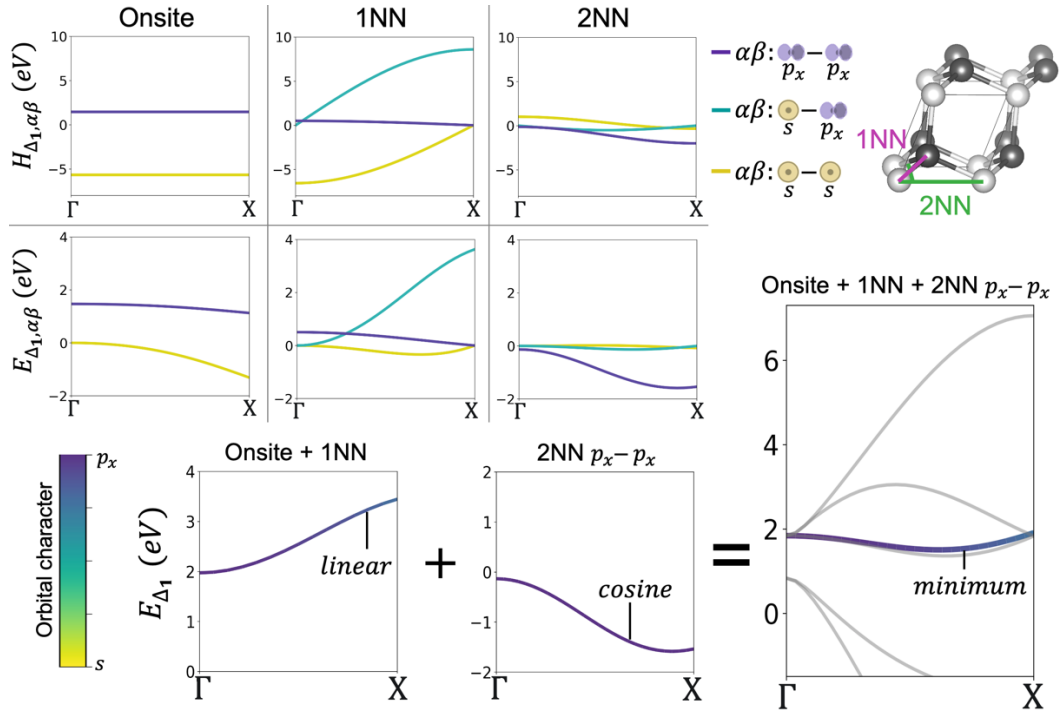

*Figure 4: Deconstructing how each bond contributes to the $\Delta_1$ conduction band in silicon by plotting the band-dependent bond runs $H_{\Delta_1,ij}$ and corresponding energies $E_{\Delta_1,ij}$ for the onsite (atomic orbital energy), 1NN, and 2NN bonds.* Interactions between $p_x$-$p_x$, $s$-$p_x$, and $s$-$s$ orbitals are colored purple, teal, and yellow, respectively, which corresponds with the orbital character color bar used in the total energy plots $E_{\Delta_1}$. Altogether, the linear behavior near X achieved with onsite + 1NN bonds and the cosine shape of the 2NN $p_x$-$p_x$ bond combine to form the minimum near X. The gray lines in the bottom right plot show the Silicon band structure with all onsite, 1NN, and 2NN bonds, where the small error between the colored and gray $\Delta_1$ band results from including the 2NN $s$-$p_x$ and $s$-$s$ bonds.

*Towards Bonding-by-Design*

Band engineering for solar cells, semiconductors, and thermoelectrics frequently requires control over the energy level of bands at specific $k$-points. Because we now have a theoretical pathway to connect the bonding interactions to the band structure, we can examine the *inverse* electronic structure design problem—*How can I modify chemical interactions to morph an existing band structure to a new band structure with more desirable features*? As a representative example, here we will modify the bonding interactions to shift the CBM from the Γ-X line in silicon to the L point as it is in germanium. This illustration paves the way to a vision of *bonding-by-design*, where instead of searching for pristine materials with a given band structure feature, we can rationally tune the chemistry (by substitutional doping or alloying) to morph a given band structure towards a desired one.

Thoroughly analyzing a band structure feature is arduous, which motivated us to create a computational analysis package (https://viz.whsunresearch.group/tb/) which systematically executes our three-step process. Our package features an interactive interface that populates tables with the orbital character and important bonds for any selected point of the $E(k)$ diagram. In addition, the band-dependent bond runs and bond energies for any of the important bonds can be plotted upon selection, allowing a user to rapidly discern how each bond contributes to the band shape. A detailed explanation and tutorial are provided in **SI.6** and **SI.7**. Our new utility marks an improvement from tools which plot the atomic orbital character on band structure[47,48] (only the first step in our 3-step process) by bringing the band energy interpretability of the Extended Hückel method[49,50] to the higher accuracy of density functional theory. Here we use our package to demonstrate how different segments of a band can be selectively raised or lowered towards a desired shape by modifying a single bond.

**Figure 5a** demonstrates a lower off-X minimum is achieved by strengthening the 2NN $p_x$-$p_x$ hopping parameters which, as described previously, contributes a shifted cosine shape to the $\Delta_1$ band. As k-point moves along the $\Gamma - L$ line in the [111] reciprocal-space direction, the orbitals of the nearest-neighbor atom in the (111) real-space direction change phase in accordance with $e^{i\mathbf{k}\cdot\mathbf{R}}$. The change in phase facilitates bonding at the L point between the $p$–$p$ orbitals of atoms with a displacement along (111). Thus, in **Figure 5b** the L point is lowered by strengthening this 1NN $p$–$p$ hopping. The lowest conduction band at Γ is an antibonding electron wavefunction of entirely $s$ orbital character. Thus, in **Figure 5c** the Γ point is lowered by weakening the 1NN $s$–$s$ hopping parameters. Notably, the $s$ character dramatically decreases from 100% to ~50% at only 0.1L along the Γ to L line, which creates the sharp curvature near Γ and leads to low effective masses in direct-gap tetrahedral semiconductors like GaAs.

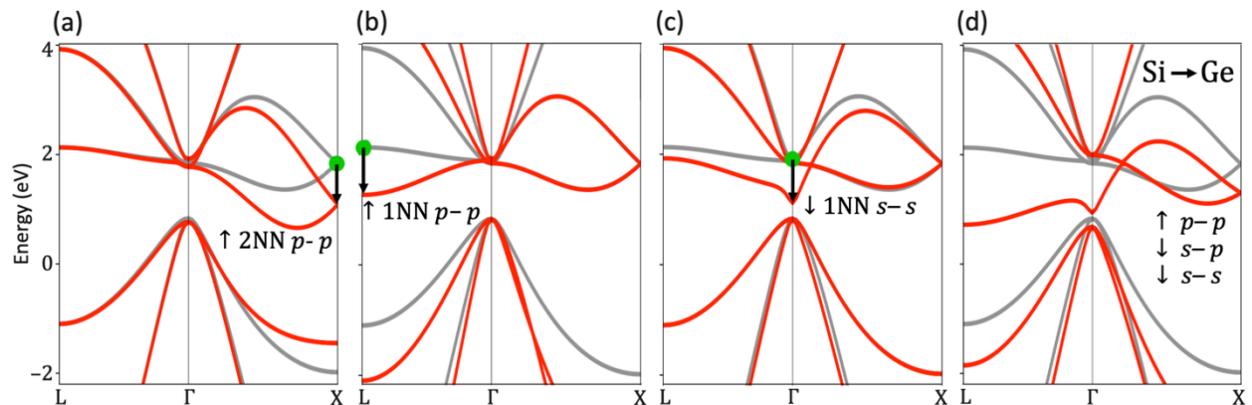

**Figure 5: The role of chemical bonding in band extrema.** In (a-d) the original Si band structure (gray) is plotted against augmented band structures (red), which changes one bond to selectively lower X, L, or Γ (plots a, b, and c, respectively) or changes bonds based on the chemistry of Ge (plot d).

Altogether, it is possible to morph the Si band structure towards the Ge band structure by increasing bonding between $p$ orbitals while decreasing bonding of $s$ orbitals. This effect is qualitatively consistent with changing chemistry from Si to Ge. In Ge, the occupied $d$ shell incompletely screens nuclear charge which attracts the valence shell—an effect called scandide contraction. The additional density near the nucleus in *s* orbitals is disproportionately impacting by the poor screening, thereby reducing the *s* orbital radius of Ge compared to the *p* orbital radii. Yuan *et al*. also found that the *d* orbitals are important in changing CBM location [51], which we find results from the indirect screening effects of *d* orbitals on *s* and *p* orbitals. Thus, when augmenting Si in **Figure 5d** by increasing all 1NN *p*–*p* interactions by 30% and decreasing all 1NN *s*–*p* and *s*–*s* bonds by 15%, we reproduce the characteristic band structure of Ge with a CBM at the L-point. Further discussion of the Si and Ge MLWF band structure can be found in **SI.8**.

The theoretical framework and automated analysis package developed in this manuscript can readily be applied to any tight-binding (TB) model. While our package does not yet directly support spin-orbit coupling, since it is not relevant in silicon, incorporating spin orbit coupling is theoretically straightforward. One would decompose the bands into bonds that specify both the orbital *and spin* of each electron, where spin-orbit coupling driven changes would manifest in the interaction between orbitals of different spin.

While our framework is generally applicable, it is limited by whether a reliable TB model can be generated from DFT. For silicon, we have justified the use of maximally localized Wannier functions (MLWFs) to generate a TB model (justification in SI.3). However, analysis using MLWF may not always yield chemically interpretable results. When the Wannier functions deviate too much from atomic orbitals—becoming combinations of multiple atomic orbitals across several atoms—our three-step process cannot elucidate the *atomic* origins of electronic structure features. Instead, it reflects the behavior of these hybridized Wannier functions. Despite these challenges, our method holds promise for application to a broader range of materials. Future work will focus on extending our framework to incorporate spin-orbit effects and developing strategies for constructing suitable TB models in complex systems.

**Outlook**

Here we presented a computable and chemically motivated framework that considers **1)** *Which orbitals are in a band*, **2)** *How are they allowed to bond*, and **3)** *How strongly do they bond?* This framework produces a sparse and therefore interpretable tight-binding model that can help us intuitively understand the crystal chemistry origins of band structure. When we applied our approach to silicon, we found that the low-symmetry conduction band minimum of silicon originates primarily from 2NN $p_x$–$p_x$ bonds, which significantly lowers the energy at X. The significance of the 2NN $p_x$-$p_x$ orbital bond compared to the 1NN is explained from the geometry of the bonding angles, in addition to there being 3× as many 2NN atoms than 1NN. This explanation is a revision on Vogl's *sps\** model, which captures the CBM position in Silicon but at the expense of other conduction bands in the Brillouin zone. Our interpretation provides a clear physical mechanism compared to previous *sp* models with multiple NN.

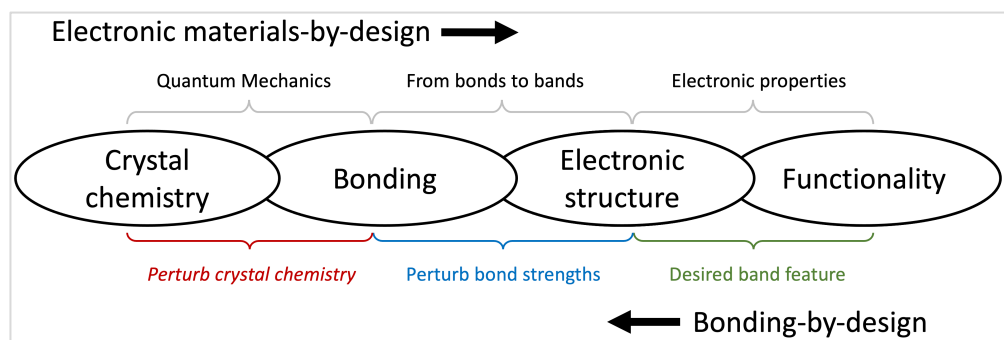

*Figure 6:* In materials-by-design, the electronic properties of pristine materials are calculated from DFT. By inverting this paradigm to bonding-by-design, one starts with the desired band feature for a given application, and rationally tunes the crystal chemistry to achieve this band feature.

Broadly speaking, our approach allows us to pinpoint the physical origin of electronic structure features in complex 3D crystals. This framework is general and can be applied to any tight-binding interpolation of a DFT-calculated band structure. By better understanding how crystal chemistry translates to major electronic structure features, we can more intuitively design chemistries and bonding environments to yield a desired band structure feature. A major advantage of this approach is the opportunity to search within the 'perturbation space' of a given material, allowing us to find best-in-class semiconductors which are often minor perturbations (strain, doping, alloying, *etc.*) from their pristine forms. This approach would invert the design paradigm from electronic 'materials-by-design' to the inverse approach of *bonding-by-design* (**Figure 6**)—where instead of searching for materials with specific properties, we can chemically or structurally modify the band structure of a given material to tune it towards next-generation electronic, optic, thermoelectric, and correlated quantum materials.